\documentclass[draft,numberedheadings]{aipproc_vw}
\layoutstyle{6x9}
\usepackage{epsfig,amsmath,natbib}

\begin{document}

\title[Environment or Outflows?]{Environment or Outflows?  New insight into the origin of narrow associated QSO absorbers}
\author{Vivienne Wild}{address={Institut d'astrophysique de Paris, 98bis Boulevard Arago,
  75014 Paris, France}
}
\date{\today}

\maketitle
%\nopagebreak
%\tableofcontents
%\cleardoublepage
\pagenumbering{arabic}

% symbols for references:       
\def \aj {AJ}
\def \mnras {MNRAS}
\def \apj {ApJ}
\def \apjs {ApJS}
\def \apjl {ApJL}
\def \aap {A\&A}
\def \nat {Nature}
\def \araa {ARAA}
\def \pasp {PASP}
\def \aaps {AAPS}

\section{Introduction}

\begin{figure}
\centering
\vspace{1cm}
\includegraphics[width=\textwidth]{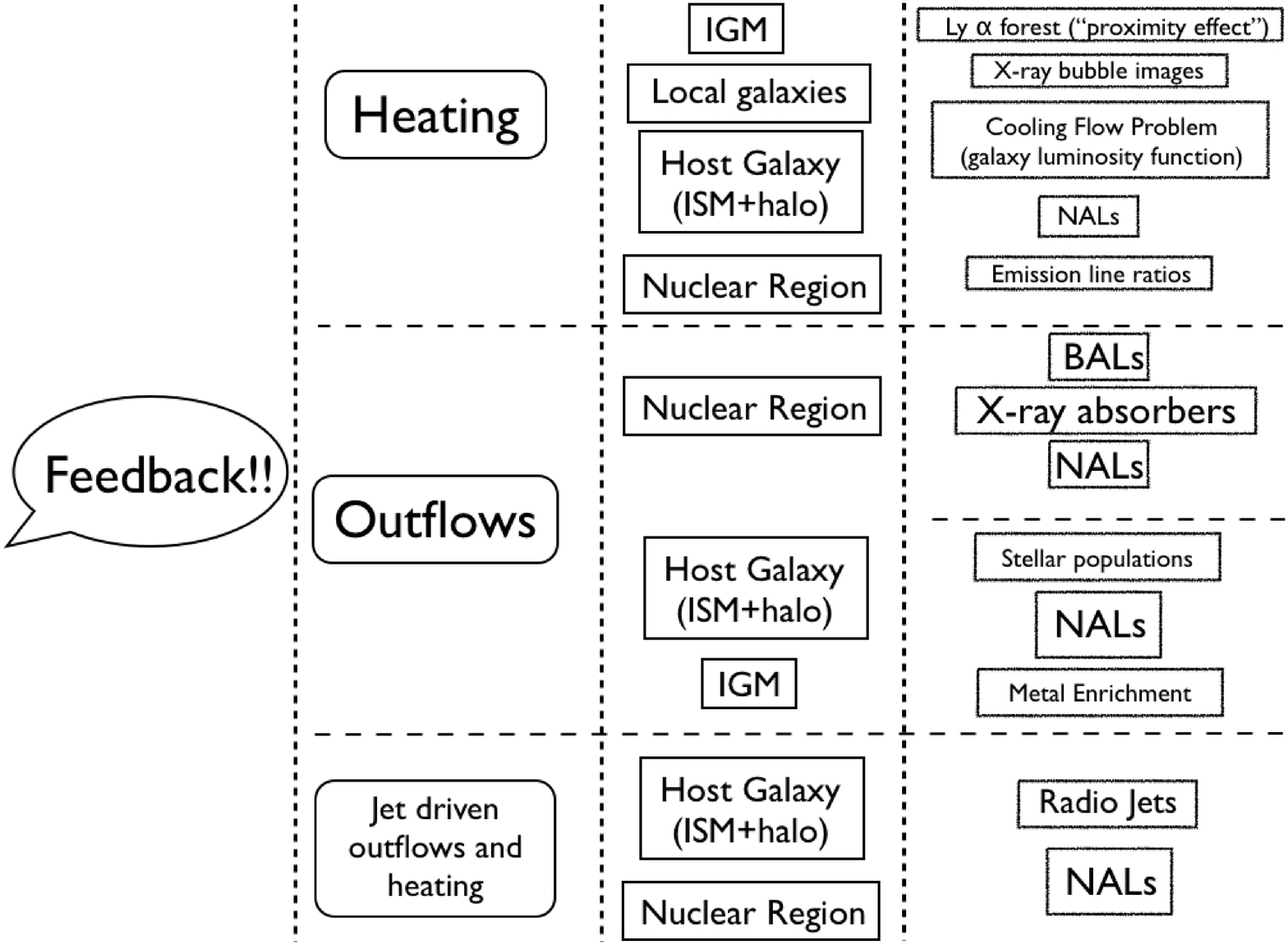}
\caption{What do we mean by AGN feedback?}
\label{fig:fb}
\end{figure}

%% All the different aspects of feedback, from observational perspective
AGN feedback is widely proposed as the solution to a number of
otherwise difficult-to-explain problems in extra-galactic
astrophysics. From an observational perspective, it is worth first
dissecting the forms of ``feedback'' that are under discussion, before
embarking on any project to observe this potentially universal
process. Figure \ref{fig:fb} gives a short summary of the topic of
feedback, which can broadly be split into two parts (column 2):
heating of gas in-situ, and outflows which remove matter from the host
galaxy. Both processes may, or may not be associated with jets, so
jets have been placed separately. While outflows are assumed to
predominantly affect the nuclear region and possibly the ISM of the
host galaxy, in-situ heating of the gas must occur on very large
scales within the IGM (column 3). The final column presents a
selection of observed or yet-to-be-observed consequences of the
physical mechanisms: the list is not meant to be exhaustive, but
simply present the range of the observations with which we must
deal. While there is little argument that some aspects of AGN feedback
have been directly detected, conclusive evidence for routine quenching
of star formation and removal of the interstellar medium of QSO host
galaxy remains elusive.

%% NALs appear everywhere in the list
Of particular relevance to this contribution, are the Narrow
Absorption Line systems (NALs) which appear in every box on the right
hand side of Figure \ref{fig:fb}, and are arguably one of the best
candidates for {\it directly} detecting ``ubiquitous'' QSO
feedback. These absorption lines, generally detected in the rest-frame
ultra-violet due to the convenient gathering of several strong
transitions, are caused by clouds of ionised and/or neutral gas which
intervene between a strong light source and the observer. This
material need not be dense to cause significant absorption of the
traversing light: the detection of MgII absorption generally implies
Hydrogen column densities of a few $10^{17}$ atoms/cm$^2$, for CIV
this number is two orders of magnitude smaller at
$\sim10^{15}$atoms/cm$^2$. In contrast to the well studied Broad
Absorption Line systems (BALs), NALs are thought in general to arise
from the interstellar medium and surrounding halo gas of ordinary
galaxies. Together with the convenient placing of their resonance
transition lines in the observed-frame optical at redshifts of
interest for QSO feedback, metal NALs can be a very powerful tool for
probing the physical state and position/velocity of gas both within
and external to the host galaxies.

%% Outline the proceedings
In these proceedings I will review some of the studies using NALs to
look for direct evidence of QSO feedback, from detailed studies of a
few objects through to statistical studies using the largest databases
of absorbers available to us today. I will present new results on the
distribution of line-of-sight velocity offsets between MgII absorbers
and their background QSOs, which reveal a high-velocity population
similar to that observed recently for CIV.

\section{Using ultra-violet NALs to reveal QSO feedback}

%% What can be used for 
Although their is an abundance of data on NALs, covering most of the
age of the Universe, and they evidently have the potential to trace
the very gas clouds we hope to see being expelled from galaxies, as
with many aspects of QSO absorption line studies real scientific
progress has been relatively slow. This can be ascribed to two main
problems. Firstly, the degeneracy between cosmological distance and
velocity makes it difficult to uniquely identify an individual
absorber with gas that is intrinsic to the host and outflowing, when
an intervening galaxy could produce the same absorption signal at the
same redshift offset. The very ubiquity of the absorbing clouds
provides a large ``contaminant'' population of intervening systems,
entirely unrelated to the problem at hand. Then there is the puzzling
absence of absorption line systems at the redshift of the QSO
\cite{Tytler:1982p1692}. This could either be due to the QSO host
being gas poor, perhaps QSOs are only observed after the expulsion of
their gas, perhaps the gas is heated to such an extent that the lines
are no longer visible, or perhaps the QSO redshifts in the small
samples studied have not been measured accurately enough to locate
$z_{ABS}\sim z_{QSO}$ systems reliably.

Detailed studies of a small number of systems have successfully shown
that some NALs are indeed intrinsic to the QSO-host
system.  Both time-variability and the presence of lines which are not
``black'' can indicate that NALs originate from within the nuclear
region. For example, Misawa et al. \cite{Misawa:2007p1867} studied a
sample of 37 high resolution QSO spectra with $2<z<4$ to conclude that
at least 50\% of quasars host high ionisation NALs (CIV, NV, SiIV)
which are diluted by unocculted light, and thus lie close to the
central engine. Detailed observations of multiple transitions have
resulted in mass loss rates and precise distances for a handful of
objects at both high and low redshift
(e.g. \cite{2007ApJ...659..250C,Rix:2007p2053}). However, with such
small samples, and especially at high redshift where imaging is
difficult in front of a bright background QSO, the question always
remains as to whether an absorption system at a few kpc is simply
the sign of an intervening galaxy. 

Detailed analyses of NALs in QSO sightlines that pass close to
foreground QSOs also allow us to probe the effect of the QSO on gas
which does not lie directly within ``firing range'' of the QSO (the
``transverse proximity effect''). Recent results remain inconclusive:
Bowen et al. \cite{2006ApJ...645L.105B} find no evidence for a
reduction in strong MgII systems, whereas Gon{\c c}alves et
al. \cite{2008ApJ...676..816G} find a significant change in ionisation
state of gas on scales of 1\,Mpc. Hennawi \& Prochaska
\cite{2007ApJ...655..735H} detect an isotropy in the distribution of
17 Lyman-limit systems around QSOs, suggesting that the line-of-sight
systems may be photoevaporated.

Such detailed analyses of small numbers of systems have been
complemented by the statistical analyses of large samples of
NALs. Until recently the question primarily revolved around the
presence, or absence, of an excess of absorbers close to QSOs (so
called ``associated'' systems, with velocities below a few hundred, to
a few thousand km/s depending on the study). With large samples, an
excess of absorbers at $z_{ABS}\sim z_{QSO}$ has now been clearly
detected (e.g. \cite{2001ApJS..133...53R,2003ApJ...599..116V}). Vanden
Berk et al. \cite{VandenBerk:2008p1707} compared the properties of
associated absorbers to those at larger redshift separations, finding
that they are dustier and have higher ionisation states. But the
ambiguity remains as to whether the population arises from
neighbouring galaxies or from gas associated with the QSO, its host
galaxy and its halo.

\section{The line-of-sight distribution of NALs in front of QSOs}

%% The SDSS catalog
The Sloan Digital Sky Survey (SDSS) has led to an enormous increase in
the quantity of data available on QSO absorption line systems. While
the spectra are not of particularly high resolution or signal-to-noise
ratio (SNR), preventing detailed analyses of individual systems, the
shear numbers of objects allow statistical studies which were
previously impossible. Here we present the topic of the statistical
analysis of associated NALs, through a new analysis of MgII absorption
line systems based on a catalog of nearly 20,000 systems culled from
the sixth data release (DR6) of the SDSS survey. The catalog was
constructed using a matched-filter detection algorithm as described in
Wild et al.  \cite{2006MNRAS.367..211W}. For reasons of catalog
completeness, the NALs are restricted to have rest-frame equivalent
width $W_{\lambda 2796}>0.5$ and the QSO spectra searched are required
to have per-pixel-SNR$>8$. For the analysis of $z_{ABS}\sim z_{QSO}$
absorption systems, accurate redshifts for the QSOs are crucial
\cite{2008MNRAS.386.2055N}, a difficult problem due to the substantial
broadening of the emission lines in QSOs. The results presented here
rely upon new QSO redshifts using a combination of available narrow
emission lines and new cross-correlation templates (Hewett \& Wild in
preparation).

\begin{figure}
\centering
\vspace{1cm}
\includegraphics[width=0.7\textwidth]{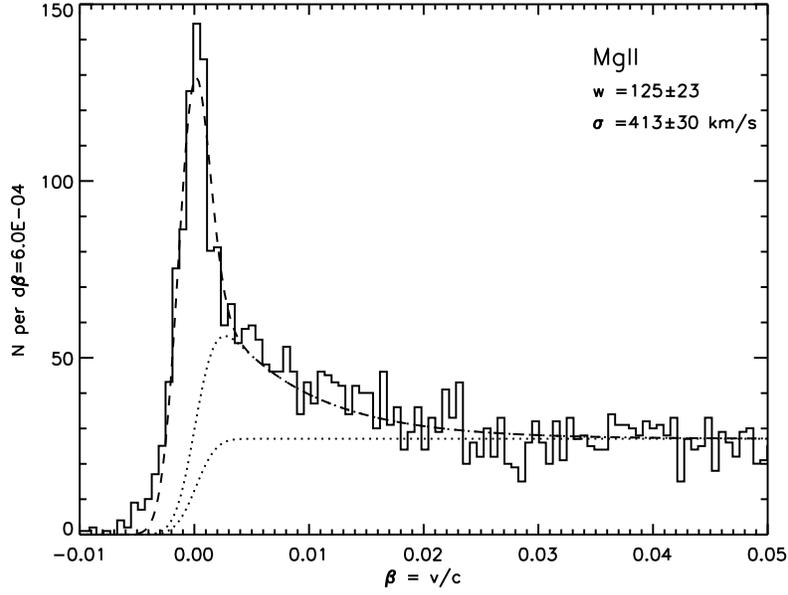}
\caption{The distribution of line-of-sight velocity offsets between
  MgII NALs and their background QSOs (histogram). Overplotted as a
  dashed line is a toy model composed of three populations:
  low-velocity absorbers; high-velocity absorbers; and intervening
  absorbers with a constant space density (see text). Dotted lines
  indicate the individual contributions from the latter two
  components. The model is convolved with a Gaussian kernel to account
  for redshift errors and/or peculiar velocities.}
\label{fig:db}
\end{figure}

In Figure \ref{fig:db} we present the distribution of velocity offsets
between the MgII absorption line systems and their background QSOs:
\begin{equation}
\beta = \frac{R^2 -1}{R^2 +1}
\hspace{0.5cm} {\rm where} \hspace{0.5cm}
R = \frac{1+z_{QSO}}{1+z_{ABS}} 
\end{equation}
For the first time for MgII, we can clearly identify three populations:
\begin{itemize}
\item At large velocities, $\beta>0.02$, the constant number density
  is consistent with an intervening population of absorbers caused by
  galaxies and gas clouds that are not physically associated with the
  QSO. 
\item A clear spike in the numbers is seen at $\beta=0$, consistent
  with a Gaussian distribution with mean of approximately zero and
  width of a few hundred km/s. Whether these NALs primarily originate
  in galaxies clustered around the QSO or in the QSO host galaxy is
  the question we must address.
\item Finally, there is a very clear extended excess of absorbers out
  to velocities $\beta<0.02$ or $v<6000$km/s, a feature previously
  seen clearly in CIV \cite{1999ApJ...513..576R,2008MNRAS.386.2055N,2008MNRAS.388..227W},
  but only hinted at before for MgII \cite{2008MNRAS.388..227W}.
\end{itemize}
The distribution is well described by a low-velocity component (delta
function) centered approximately on zero, an exponentially distributed
high-velocity component of width $w$ at $\beta>0$ (upper dotted line),
both superposed on a constant background intervening population ($B$,
lower dotted line), and convolved with a Gaussian kernel to account
for redshift errors and/or peculiar velocities:
\begin{equation}
N_{ABS} = \left( A_1\delta(\beta-\mu) + \left[A_2 \exp{(w \beta)}
  + B\right]_{\beta>0} \right) * G(\sigma)
\end{equation}
where we find $\sigma=413\pm30$km/s and $w=125\pm23$. The full fit
is shown as a dashed line in Figure \ref{fig:db}.

Unfortunately the detection of an excess of NALs at the redshift of
the QSO is not unambiguous evidence for NALs in the host galaxies of
the QSOs. As we show in the next section, galaxy clustering can lead
to a signal which is difficult to distinguish with current data.

%%%%%%%%%%%%%%%%%%%%%%%%%%%%%%%%%%%%%%%%%%%%%%%%%%%%%%%%%%%%%%%%%%

\section{The 3D distribution of NALs around QSOs}

\begin{figure}
\centering
\vspace{1cm}
\includegraphics[width=0.7\textwidth]{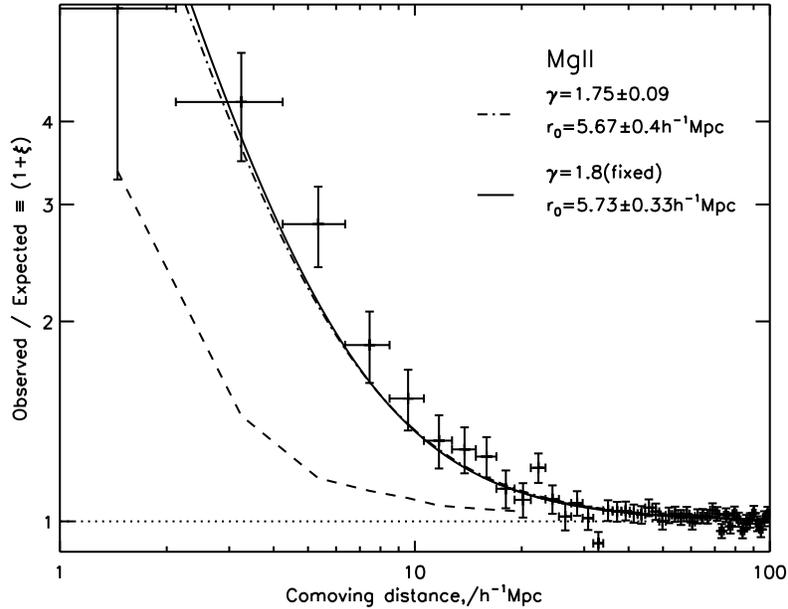}
\caption{The 3D clustering of MgII NALs around QSOs as a function of
  comoving separation.  The dashed line is the 68 per cent detection
  threshold given the number of sightlines and number density of
  absorbers. The dash-dot and solid lines are best-fit power laws with
  parameters given in the top right.}
\label{fig:trans}
\end{figure}

%% line-of-sight vs. transverse
With the size of catalogs now available from the SDSS, it is possible
to measure directly the 3D clustering of absorbers around QSOs using a
cross-correlation style analysis. With the clustering amplitude in
hand, we will then estimate the excess number of absorbers expected
along the line-of-sight to the QSOs. The method used to measure the 3D
clustering is presented in detail in \cite{2008MNRAS.388..227W}. To
summarise, we count the number of observed QSO-MgII pairs ($N_{obs}$)
as a function of comoving separation ($r$) and compare this to the
number of pairs that would expect ($N_{exp}$) for a constant
background distribution of absorbers without clustering. To avoid
contamination from NALs that might be associated with outflowing gas
from a QSO host, we restrict the NAL sample to those with
$z_{ABS}<z_{QSO}-0.1$. In Figure \ref{fig:trans} we present the new
results using the DR6 MgII catalog with improved QSO redshifts. The
dash-dot line is a powerlaw fit of the form:
\begin{equation}
\xi(r) = \frac{N_{obs}}{N_{exp}} -1 = (r/r_0)^{-\gamma}
\end{equation}
where $r_0$ is the correlation scale length which we measure to be
$5.67\pm0.4h^{-1}$Mpc, with a power law index of $\gamma=1.74\pm
0.09$, or $5.73\pm0.3h^{-1}$Mpc at fixed $\gamma=1.8$. This
correlation length is similar to that measured for bright galaxies at
similar redshifts. There is evidence, at the level of around
$3\sigma$, for a flattening in the MgII-QSO clustering on small scales
($<5h^{-1}Mpc$). This may be caused by QSO redshift errors or absorber
peculiar velocities which can have a significant effect at small
absorber-QSO separations. It may also indicate the presence of a
transverse ``proximity effect'', where the QSO ionises the gas in its
surrounding halo (but see \cite{2006ApJ...645L.105B}). Clearly there
is scope for further investigation of this feature in the
future. Discarding the central bin from our power law fit increases
the measured correlation length and power law index both by about
$3\sigma$, leading to a larger predicted clustering signal that only
enhances the qualitative conclusions drawn from this study.

\section{The clustering contribution to the line-of-sight
  excess}
\begin{figure}
\centering
\vspace{1cm}
\includegraphics[width=0.7\textwidth]{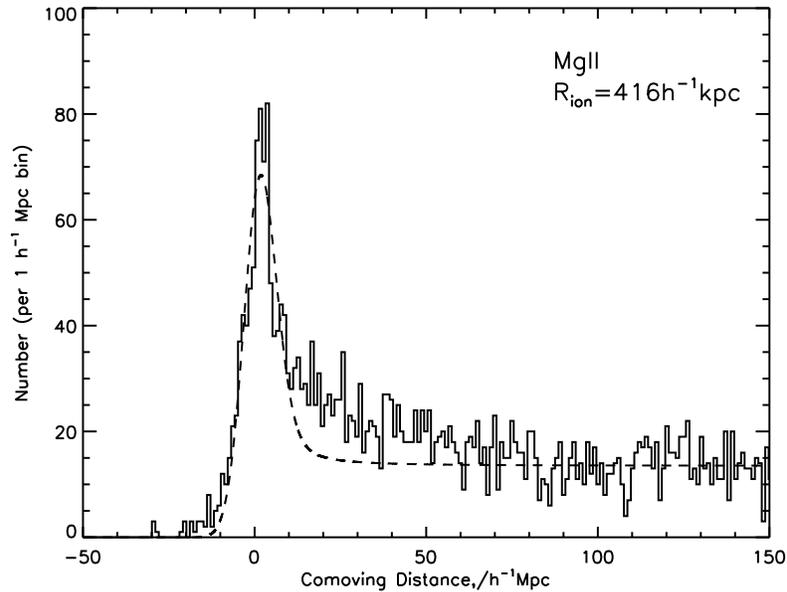}
\caption{Distribution of line-of-sight QSO-absorber separation as a
  function of comoving distance (where line-of-sight redshift
  separation has been converted into comoving distance in the usual
  way). The dashed line shows the predicted distribution of MgII
  absorbers from clustering, assuming the QSO ionises (i.e. removes)
  all absorbers to a proper distance of $130h^{-1}$kpc.}
\label{fig:dr}
\end{figure}

In Figure \ref{fig:dr} we convert the distribution of line-of-sight
QSO-absorber redshift separations into comoving distance units, to
allow direct comparison with the 3D-clustering results of the previous
section. One free parameter is then required for our toy-model: the
distance below which physical processes internal to the host-galaxy
dominate the distribution of absorbers, rather than the clustering of
galaxies in the QSO neighbourhood. As we shall see, the precise
distribution of MgII absorbers around the host galaxy of the QSOs is
irrelevant to our results, due to the significant deficit of absorbers
detected along the line-of-sight. We therefore define a simple
``ionisation radius'' (R$_{\rm ion}$) internal to which the number of
absorbers is zero. Finally, the model is convolved with a Gaussian of
width equivalent to $\sigma=413$km/s at the median redshift of
absorbers which lie within $\pm10h^{-1}$Mpc of their background QSO
($z=1.3$), i.e.  to match the measured width of the distribution in
velocity space (Figure \ref{fig:db}). The dashed line in Figure
\ref{fig:dr} shows the predicted line-of-sight distribution of
absorbers in front of QSOs, from galaxy clustering alone and with a
ionisation radius $\sim420h^{-1}$kpc (comoving units), or
$180h^{-1}$kpc (proper units at the median redshift of the sample). We
note that this value for the ionisation radius is slightly lower than
that given in \cite{2008MNRAS.388..227W}, likely resulting from the
completely independent method used to select the absorbers.

Typical MgII halos around galaxies can extend to $\sim40h^{-1}$kpc
(proper) with almost unity covering fractions
\cite{1993eeg..conf..263S}. Beyond this distance MgII halos are
thought to be patchy, and can extend to distances of $70h^{-1}$kpc
\cite{2008AJ....135..922K,2007ApJ...658..161Z}. Our result shows that
the MgII ion, with an ionisation potential of 15.03eV, is
destroyed in clouds which lie at even greater distances from QSOs and
thus far into the IGM. The spike of absorbers below $\beta<0.002$,
$v<600$km/s, or $R<10h^{-1}$Mpc, is entirely consistent with galaxy
clustering from galaxies that lie beyond the ionisation zone of the
QSO. However, we cannot rule out that R$_{\rm ion}$ is indeed even
larger and the low-velocity absorbers are caused by denser,
self-shielded, clouds remaining within the ionisation zone, perhaps
even intrinsic to the QSO host itself.

Clearly the high velocity tail is, however, caused by a process
internal to the QSO itself, and with velocities as high as $0.02c$
($\sim6000$km/s) these outflows must be driven by the central AGN
engine, rather than any accompanying starburst
\cite{2007ApJ...663L..77T}. The very existence of these absorbers is
puzzling, given the clear ability of the QSO radiation to destroy all
normal MgII clouds out to very large distances. Their existence also
leads us to question the conclusion that the low-velocity absorbers
are primarily due to galaxy clustering. Are they the remnants of the
densest ISM clouds yet to be destroyed? Are they unrelated to ordinary
MgII ISM clouds, and instead created in the turbulence of outflowing
gas?  Their distance from the nuclear source remains to
be determined. If they are external to the nuclear region, then they
are surely evidence for the expulsion of (cold) gas from the galaxy
ISM. If they are internal to the nuclear region, such low ionisation
gas with narrow velocity widths can constrain models for the inner
regions of QSOs \cite{2000ApJ...545...63E}.

%From comparison to detailed studies of small numbers of
%QSOs at similar redshifts, about 50\% 
%distance of these absorbers from the central engine remains to be
%discovered, but their
%and leads to the obvious question of how
%these MgII absorbers are existing in the intense radiation field of
%the QSO, that is clearly destroying all

\section{Radio Loud vs. Radio Quiet}
One further statistical investigation may lead to significant insight
into the origin of the MgII absorbers within 6000km/s from the QSO. It
has been known for some time that QSOs with different radio properties
(loud/quiet, flat/steep spectrum) have different fractions of
absorption line systems, strongly suggesting a non-intervening origin
at least for a subset
\cite{1994ApJS...93....1A,2001ApJS..133...53R}. More recently,
detailed studies of nearby radio galaxies have revealed outflowing
neutral Hydrogen in 21cm absorption against the background radio
source \cite{2005A&A...444L...9M}.

\begin{figure}
\centering
\vspace{1cm}
\includegraphics[width=0.7\textwidth]{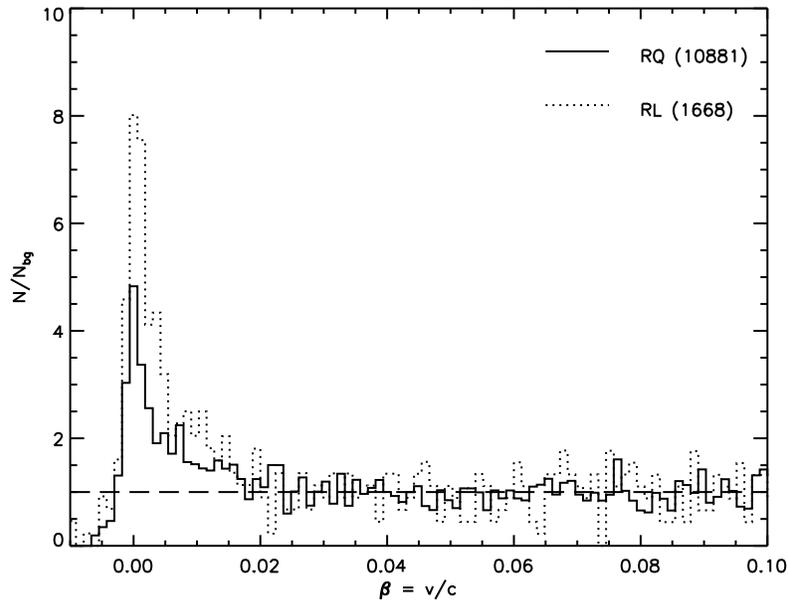}
\caption{The distribution of line-of-sight velocity offsets between
  MgII NALs and the background QSOs, separated by the radio luminosity
  at $10^{25}$W/Hz of the QSOs. The RQQSO sample has been selected to
  match the RLQSO sample in optical luminosity.}
\label{fig:radio}
\end{figure}

In Figure \ref{fig:radio} we present the velocity separation of radio
loud QSOs (RLQSOs, $L_{\rm FIRST} > 10^{25}$W/Hz) compared to a sample
of radio quiet QSOs (RQQSOs) matched in optical luminosity to the
RLQSOs. We can clearly see that RLQSOs show a larger excess of
low-velocity MgII absorbers than RQQSOs with high significance. Within
$-0.002<\beta<0.002$, RLQSOs have an excess of 6.2 absorbers over the
background level, compared to 3.6 for RQQSOs. At high velocities,
RLQSOs also seem to show a small increase in MgII NALs: for
$0.002<\beta<0.02$, RLQSOs have an excess of 2.1 compared to 1.6 for
RQQSOs.

The question remains as to whether RLQSOs are more strongly clustered
than RQQSOs.  If so, then the excess low-velocity MgII absorbers seen
in RLQSOs may result solely from them living in higher density
neighbourhoods. Unfortunately, the SDSS DR6 catalog is still not quite
large enough to answer this question using the method presented above.

\section{Conclusions}

A number of recent studies have found that NALs intrinsic to the QSO
host can be found in at least 50\% of QSO spectra, and in most cases
they are found to be outflowing. Simply due to observational
limitations the position of these confirmed cases is, however, close
to the central nucleus. In the few cases where larger distances can be
determined from detailed line analyses, it is usually impossible to
rule out the presence of an intervening galaxy. 

Through the enormous statistical power of the SDSS, we can now
determine precisely the contribution of galaxy clustering to
QSO-absorber line-of-sight distributions. This leads to the
following conclusions:
\begin{itemize}
\item QSOs heat the gas to considerable distances along their
  line-of-sight, with relatively low ionisation MgII ions ionised to
  several hundred kpc (comoving) into the IGM.
\item Within 600km/s there is an excess of NALs, however, this excess
  is most simply explained by ordinary absorption clouds in and around
  galaxies which lie outside of the ionising influence of the QSO.
\item A subset of absorbers out to velocities of 6000km/s (MgII) or
  12000km/s (CIV) can not be explained by intervening galaxies. Their
  velocity distribution is well fit by a declining exponential (but
  see \cite{2008MNRAS.386.2055N}), and their high maximum velocities
  indicate an origin close to the central engine.
\item There is a significant excess of low-velocity NALs in RLQSOs,
  compared to RQQSOs. This excess may also extend into the
  high-velocity systems. Unfortunately, the statistics are not quite
  good enough to rule out the possibility that RLQSOs
  simply live in denser environments.
\end{itemize}
The heating effect of a QSO on its host galaxy, and likewise on all
nearby galaxies, is unmistakable.  However, the existence of the
high-velocity systems, which we would naively expect not to exist in
the intense radiation field of the QSO, leaves a narrow window of
doubt as to the true origin of the low-velocity systems. Allowing the
ionisation radius to increase, thus removing more intervening clouds,
would allow some, if not all, low-velocity systems to arise from gas
associated with the QSO, its host galaxy and its halo. There is
certainly more work to be done before we can definitively
claim the origin of low-velocity NALs to be intervening galaxies.

\section{Acknowledgements}
I would like to thank all of the team that helped me complete the
first stage of this work, Paul Hewett for the new catalog and QSO
redshifts, the conference participants for their enthusiastic
discussions and the conference organisers for putting together such an
interesting program.

Funding for the SDSS and SDSS-II has been provided by the Alfred
P. Sloan Foundation, the Participating Institutions, the National
Science Foundation, the U.S. Department of Energy, the National
Aeronautics and Space Administration, the Japanese Monbukagakusho, the
Max Planck Society, and the Higher Education Funding Council for
England. The SDSS Web Site is http://www.sdss.org/.

% \begin{thereferences}{99}

%  \label{reflist}
%   \bibitem{abbott}
%   Abbott, L.F. and Deser, S. (1982). Stability of gravity with a
%   cosmological constant, \textit{Nucl. Phys.} \textbf{B195}, 76--96.

%   \bibitem{adams}
%   Adams, J.F. (1981). Spin (8), triality, $F_4$ and all that, in
%   \textit{Superspace and Supergravity}, ed. S.W.~Hawking and M.~R\"ocek
%   (Cambridge University Press, Cambridge).

%   \bibitem{arnold}
%   Arnol'd, V.I. (1978). \textit{Mathematical Methods of Classical
%   Mechanics} (Springer, New York).

%   \bibitem{buch}
%   Buchdahl, N.P. (1982). Applications of Several Complex Variables to
%   Twistor Theory, Oxford University D. Phil. thesis.

% \end{thereferences}

%\cleardoublepage

\end{document}